\documentclass[12pt]{iopart}
\usepackage[caption=false]{subfig}
\usepackage{tabularx}
\usepackage{graphicx}
\graphicspath{{./figs/},{.}}
\usepackage{dcolumn}
\usepackage{bm}
\usepackage{float}
\usepackage[Gray,squaren,thinqspace,thinspace]{SIunits} 
\usepackage{amssymb}
\usepackage{ifthen}
\usepackage{xspace}
\newcommand{\ket}[1]{\mid#1\rangle}
\newcommand{\bra}[1]{\langle#1\mid}

\newcommand{\braket}[2]{\langle#1\mid#2\rangle}
\newcommand{\G}{\ensuremath{\widehat{\mathcal{G}}}}
\newcommand{\Om}{\ensuremath{\widehat{\Omega}}}
\newcommand{\Omo}{\ensuremath{\widehat{\Omega}^{[1]}}}

\newcommand{\Omeff}{\ensuremath{\widehat{\Omega}^{\text{eff}}}}
\newcommand{\Omloc}{\ensuremath{\widehat{\Omega}^{\text{LCA}}}}

\newcommand{\eq}[2][]{\ifthenelse{\equal{#1}{}}{\begin{equation}#2\end{equation}}{\begin{equation}\label{eq:#1}#2\end{equation}}}
\newcommand{\eqa}[2][]{\ifthenelse{\equal{#1}{}}{\begin{equation}#2\end{equation}}{\begin{equation}\label{eq:#1}#2\end{equation}}}
\newcommand{\eqmultline}[2][]{\ifthenelse{\equal{#1}{}}{\begin{equation}#2\end{equation}}{\begin{equation}\label{eq:#1}#2\end{equation}}}

\newcommand{\vecp}{\ensuremath{\vec{p}}}

\newcommand{\invfm}{fm$^{-1}$\xspace}

\newcommand{\eqn}[1]{~(\ref{eq:#1})\xspace}
\begin{document}

\title{Stylized features of single-nucleon momentum distributions}
\author{Jan Ryckebusch, Maarten Vanhalst, Wim Cosyn}
\address{Department of Physics and Astronomy,\\
 Ghent University, Proeftuinstraat 86, B-9000 Gent, Belgium}
\eads{Jan.Ryckebusch@UGent.be}
%
%
\date{\today}
\begin{abstract}
Nuclear short-range correlations (SRC) typically manifest themselves
in the tail parts of the single-nucleon momentum distributions. We
propose an approximate practical method for computing those SRC
contributions to the high-momentum parts. The framework adopted in
this work is applicable throughout the nuclear mass table and corrects
mean-field models for central, spin-isospin and tensor correlations by
shifting the complexity induced by the SRC from the wave functions to
the operators. It is argued that the expansion of these modified
operators can be truncated to a low order. The proposed model can
generate the SRC-related high-momentum tail of the single-nucleon
momentum distribution.  These are dominated by correlation operators
acting on mean-field pairs with vanishing relative radial and
angular-momentum quantum numbers.  The proposed method explains the
dominant role of proton-neutron pairs in generating the SRC and
accounts for the magnitude and mass dependence of SRC as probed in
inclusive electron scattering. It also provides predictions for the
ratio of the amount of correlated proton-proton to proton-neutron
pairs which are in line with the observations. In asymmetric nuclei,
the correlations make the average kinetic energy for the minority
nucleons larger than for the majority nucleons.

\vspace{0.8cm}
\noindent(Some figures may appear in colour only in the online journal)

\end{abstract}

\vspace{0.8cm}

\submitto{\JPG}
\maketitle
\section{Introduction}

Momentum distributions contain all the information about the momentum
decomposition of the nuclear ground-state wave function.  The
computation of single-nucleon momentum distributions has reached a
very high level of sophistication to date. Ab-initio methods with
variational wave functions can be used to compute the momentum
distributions for nuclei up to atomic mass number $A=12$
\cite{Schiavilla:2006xx,Wiringa:2008vv,Feldmeier:2011tt,PhysRevC.89.024305,Alvioli:2013ff}.
Also for atomic mass number infinity, or nuclear matter, advanced many-body 
calculations with realistic nucleon-nucleon interactions can be
performed \cite{Benhar:1989aw,Rios:2013zqa}. Momentum distributions
for mid-heavy and heavy nuclei cannot be computed with ab-initio
methods to date.  Advanced approximate schemes like cluster expansions
\cite{Alvioli:2013ff,Alvioli:2007zz,Alvioli:2012dd} and correlated
basis function theory \cite{AriasdeSaavedra:2007qg,Bisconti:2007dd}
provide momentum distributions for heavier nuclei.

Since the dawn of nuclear physics, the mean-field model has been put
forward as a good starting point for understanding the complexity of
nuclear dynamics. Important corrections to the mean-field model stem
from long-range correlations and SRC
\cite{Dickhoff:2004xx}. Long-range correlations (LRC) rather affect
the low-momentum (infrared) behaviour of the nuclear dynamics, whereas
SRC are mostly connected with the high-momentum (ultraviolet)
behaviour. As a consequence, by tuning the spatial resolution of the
probe used to study nuclei, one can reasonably separate the long-range
and short-range phenomena. The focus of this work is on the study of
SRC, and LRC are neglected. We wish to put forward
a comprehensive theoretical framework to interpret the results of the
recent measurements probing SRC, which include studies of the mass and
isospin dependence of the SRC. To this end, we present an approximate
practical way of computing the SRC contributions to momentum
distributions for stable nuclei over the entire mass range.

We start from ground-state wave functions that can be written as correlation
operators acting on a single Slater determinant. The computation of
expectation values of one-body and two-body operators for those wave
functions involves multi-body effective operators and a truncation
scheme is in order.  We propose a low-order correlation operator
approximation, dubbed LCA, that truncates the modified correlated
operator corresponding with an one-body operator to the level of
two-body operators. The LCA method is specifically designed for
dealing with correlations which extend over relatively short
distances. For the computation of the single-nucleon momentum
distribution, the LCA model developed in section~\ref{sec:formalism},
preserves the normalisation conditions.

In section~\ref{sec:obmd}, we illustrate that the LCA method is a
practical approximate way of computing the effect of SRC on
single-nucleon momentum distributions for nuclei over the entire mass
range.  It will be shown that after inclusion of central, spin-isospin
and tensor correlations, it can capture some stylized features of
nuclear momentum distributions.  Due to its wide range of
applicability, the LCA framework allows one to study the mass and
isospin dependence of SRC and to arrive at a comprehensive picture of
the impact of SRC throughout the mass table.  We compare the LCA
predictions for the high-momentum parts of the single-nucleon momentum
distributions for $^{4}$He, $^{9}$Be and $^{12}$C with those from
ab-initio calculations.

Of course, the LCA approximate method is only justified if the
resulting physical quantities like radii and kinetic energies are in
reasonable agreement with data and results from more realistic
approaches.  The impact of SRC on the average nucleon
kinetic energies and the rms radii for symmetric and asymmetric nuclei
is discussed in section~\ref{sec:kinene}.  As the correlations induce
high-momentum components, they increase the average kinetic energies.
The isospin dependence of the SRC is at the origin of some interesting
features which depend on the asymmetry of nuclei
\cite{Rios:2013zqa,Konrad:2005qm,Hen:2014,Sargsian:2012sm}. Also these asymmetry effects will
be discussed in section~\ref{sec:kinene}.

\section{Formalism}
\label{sec:formalism}
A time-honoured method to account for correlations in independent
particle models (IPM) is to shift the complexity induced by the
correlations from the wave functions to the operators
\cite{Feldmeier:2011tt,PhysRevC.86.064304}.  The correlated ground-state wave
function $\ket{\Psi}$ is constructed by applying a many-body
correlation operator $\G$ to the uncorrelated single Slater determinant 
$\ket{\Phi}$.  The operator $\G$ considered in this work, corrects the
IPM Slater determinant $\ket{\Phi}$ for SRC: \eq[defpsi]{ \ket{\Psi} =
  \frac{1}{\sqrt{\mathcal{N}}} \G \ket{\Phi}, } with the normalisation
factor $\mathcal{N} \equiv \bra{\Phi} \G^\dagger \G \ket{\Phi}$.
Determining the operator $\G$ represents a major challenge
\cite{Neff:2002nu}.  One can be guided, however, by the knowledge of
the basic features of the nucleon-nucleon force.  As far as the
short-range nucleon-nucleon ($NN$) correlations are concerned, $\G$ is
dominated by the central, spin-isospin and tensor correlations
\cite{PhysRevC.89.024305,janssen00,ryckebusch97}
\begin{equation}
\widehat{\mathcal{G}}   \approx   \widehat {{\cal S}}  
\left( \prod _{i<j=1} ^{A} \left[ 1 + \hat{l} \left(i,j\right) \right] \right)  \; ,
\label{eq:coroperator}
\end{equation}
with $ \widehat
{{\cal S}} $ the symmetrisation operator and  
\begin{eqnarray}
\hat{l} \left(i,j\right)
& = &- \hat{g}(i,j) + \hat{s}(i,j) + \hat{t}(i,j) \nonumber \\
& = & - g_c(r_{ij}) + 
f_{\sigma \tau}(r_{ij}) \vec{\sigma}_i \cdot \vec{\sigma}_j 
\vec{\tau}_i \cdot \vec{\tau}_j + 
f_{t\tau}(r_{ij}) \widehat{S}_{ij} \vec{\tau}_i \cdot 
  \vec{\tau}_j \; .
\label{eq:sumofcorrelators}
\end{eqnarray}
Here, ${\widehat{S}_{ij}}$ is the tensor operator and $r_{ij} =
\left|\vec{r}_{i}-\vec{r}_{j}\right|$. Further, $g_c(r_{12})$,
$f_{\sigma \tau}(r_{12})$ and $f_{t\tau}(r_{12})$ are the central,
spin-isospin and tensor correlation functions. The $g_c(r_{12})$
encodes the fact that nucleons have a finite size and forcefully repel
each other at short internucleon distances. There is a very strong
model dependence in the theoretical predictions for $g_c(r_{12})$
\cite{janssen00,Blomqvist:1998gq}. Predictions range from rather
``soft'' $g_c(r_{12})$ (with $\lim _{r_{12} \to 0} g_c(r_{12}) \ne 1$)
to ``hard'' ones (with $\lim _{r_{12} \to 0} g_c(r_{12}) = 1$) which
possess an exclusion zone in the short-distance radial distribution of
nucleon pairs. From an analysis of the relative pair momentum
distributions in $^{12}$C$(e,e'pp)$ experiments
\cite{Blomqvist:1998gq} one could deduce that the ``hard'' correlation
functions, like the one predicted in the G-matrix calculations by
Gearhart~\cite{gearheart94}, provide a fair account of the data.
Throughout this work we use the $g_c(r_{12})$
of~\cite{gearheart94}. The spin-isospin and tensor correlation
functions $f_{\sigma \tau}(r_{12})$ and $f_{t\tau}(r_{12})$ extend to
larger internucleon distances than $g_c(r_{12})$
\cite{Vanhalst:2012ur}. We use the $f_{\sigma \tau}(r_{12})$ and
$f_{t\tau}(r_{12})$ from the variational Monte-Carlo calculations by Pieper et al.~\cite{Pieper:1992gr}. Note that
the $g_c(r_{12})$ of~\cite{Pieper:1992gr} is very soft and severely
underestimates the relative-momentum distributions of the
$^{12}$C$(e,e'pp)$ measurements of~\cite{Blomqvist:1998gq}. The
combination of the three correlation functions considered in this
work, has also been used in theory-experiment comparisons for
semi-exclusive $A(e,e'p)$ \cite{janssen00,Fissum:2004we} and exclusive
$^{16}$O$(e,e'pp)$ \cite{Ryckebusch:2003tu}.


Evaluating the expectation value of an operator $\Om$ between
correlated states of (\ref{eq:defpsi}) is far from trivial. The procedure detailed in
 \cite{PhysRevC.86.064304} for example, amounts to rewriting the
matrix element between correlated states \eq{ \bra{ \Psi } \Om
  \ket{\Psi} , } as a matrix element between uncorrelated states
\eq[defeff]{ \frac{1}{\mathcal{N}} \bra{ \Phi } \Om^{\text{eff}}
  \ket{\Phi} \; .  } Hereby, one introduces an effective transition operator
$\Omeff$ that corrects the operator $\Om$ for the SRC effects 
\begin{eqnarray}
 \Omeff & = {} & \G^\dagger \; \Om \; \G
  \nonumber \\ & = {} & \Bigl( \prod_{i<j=1}^A \bigl[ 1-\hat{l}(i,j) 
  \bigr] \Bigr)^\dagger 
   \; \Om  \; 
  \Bigl( \prod_{k<l=1}^A \bigl[ 1-\hat{l}(k,l) 
  \bigr] \Bigr) \; .
\label{eq:omefffull}
\end{eqnarray}
For the sake of computing single-nucleon momentum distributions, it
suffices to consider one-body operators \eq[obo]{ \Om \equiv
  \sum_{i=1}^A \Om^{[1]}(i) \; .  } In the LCA framework used in this
work, a perturbation expansion for~(\ref{eq:omefffull})
is adopted.  Thereby, the local dynamical origin of the SRC is
exploited to truncate the expansion \cite{PhysRevC.86.064304,Braaten2008}.  Studies of the
single-nucleon spectral function in nuclear matter
\cite{Benhar:1989aw} reveal that the correlated part is mainly
furnished by three-body breakup processes. For a finite nucleus $A$
this translates into processes with two close-proximity correlated
nucleons and a spectator residual $A-2$ core.  This picture has been
confirmed in semi-exclusive $A(e,e'p)$ measurements
\cite{Fissum:2004we, Iodice:2007mn}.  These observations allow one to
treat the SRC as pair correlations to a good approximation. It also
justifies a perturbation expansion of~(\ref{eq:omefffull}) that truncates the effective operators $\Omeff$
corresponding with a one-body operator $ \sum_{i=1}^A \Om^{[1]} (i)$ to the level of
two-body operators. We retain the terms that are linear
and quadratic in the correlation operator $\hat{l}$. The quadratic
terms contain terms with both correlation operators acting on the same
particle pair.  This results in the following effective operator
\eqmultline[omloc]{ \Omeff \approx \Omloc = \sum_{i=1}^A \Om^{[1]}(i)
  + \sum_{i<j=1}^A \left\{ \Om^{[1],\text{l}}(i,j) + \left[
    \Om^{[1],\text{l}}(i,j) \right] ^{\dagger} +
  \Om^{[1],\text{q}}(i,j) \right\} . } Here, the linear (l) and
quadratic (q) terms read \eq[omoin]{
  \widehat{\Omega}^{[1],\text{l}}(i,j) =
  \left[ \Omega^{[1]}(i) + \Omega^{[1]}(j)  \right] \hat{l}(i,j) ,
}
\eq[omoqu]{
  \widehat{\Omega}^{[1],\text{q}}(i,j) = 
  \hat{l}^\dagger(i,j)
  \bigl[ \Omo(i) + \Omo(j) \bigr]
  \hat{l}(i,j) .
}
The LCA effective operator of \eqn{omloc} has one- and 
two-body terms, and can be conveniently rewritten  
as $\Omloc = \sum_{i<j}^A \Omloc(i,j)$ with
\eqa[omeff2]{
  \Omloc(i,j) =  
\frac{1}{A-1}\left[ \Omo(i)+\Omo(j) \right]
  + \widehat{\Omega}^{[1],\text{corr}}(i,j) \; ,
}
whereby we have introduced a short-hand notation for that part of the
operator associated with the correlations
\begin{eqnarray}
  \widehat{\Omega}^{[1],\text{corr}}(i,j) & = & 
  \widehat{\Omega}^{[1],\text{l}}(i,j)
  + \left[ \widehat{\Omega}^{[1],\text{l}}(i,j) \right] ^ {\dagger}
  + \widehat{\Omega}^{[1],\text{q}}(i,j) \; . \nonumber \\
& &
\label{eq:correlatedpartofomega}
\end{eqnarray}
In the absence of correlations only the first term in the expansion of
\eqn{omloc} does not vanish. At medium internucleon distances ($r_{ij}
\gtrsim 3 $~fm) one has that  $\hat{l}(i,j) \rightarrow 0$ and the effective
operator $\Omloc$ equals the uncorrelated operator $\Om$. The applicability of the LCA method, which involves a truncation
  of the effective operators to terms which are linear and quadratic
  in the correlation operators, hinges on the local character of the
  SRC. Long-range correlations, for example, would require an
  expansion which involves higher-order contributions.  

\begin{figure}
  \begin{center}
  \includegraphics[viewport=129 589 556 683,clip,width=\textwidth]{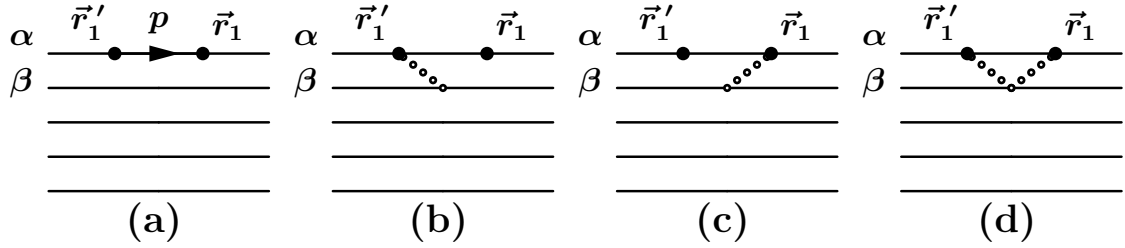}
  \end{center}
  \caption{
Diagrams (a)-(d) denote
    the different contributions to the $n^{[1]}(p)$ as it is computed
    in the LCA.  The solid lines denote nucleons
    in the single-particle state with IPM quantum numbers $\alpha, \beta,
    \ldots$ and the dotted lines are the correlation operators
    $\hat{l}$.  Diagram (a) is the IPM contribution to
    $n^{[1]}(p)$. The other diagrams are the SRC corrections. In the
    LCA we consider the diagrams that involve two nucleons and are
    either linear ((b) and (c)) or quadratic (d) in the correlation
    operators.}
    \label{fig:diagrams}
 \end{figure}

The single-nucleon momentum distribution $n^{[1]}(p)$ quantifies the
probability of removing from the nuclear ground state a momentum $p$
at $\vec{r}^{\; \prime}_1 $ and putting it instantly back at
$ \vec{r}_1 $ for any combination of $ \left( \vec{r}_1 , \vec{r}^{\, \prime}_1 \right)$.  Accordingly, $n^{[1]}(p)$ is connected to
the expectation value of the operator $ \widehat{\psi} ^ {\dagger}
\left( \vec{r}_1 \right) \widehat{\psi} \left( \vec{r}_1 ^{\; \prime}
\right) $ (the nucleon field operator $ \widehat{\psi} \left(
\vec{r}_1 ^{\; \prime} \right)$ annihilates a nucleon at position $\vec{r}_1 ^{\; \prime}$) in the exact ground state
$\Psi$. One can write $\left( d^{3(A-1)} \{ \vec{r}_{2-A} \} \equiv
\prod _{i=2} ^{i=A} d^{3}\vec{r}_i \right)$ 
\begin{equation}
\fl     
n^{[1]}(p) = {} \int
  \frac{d^2 \Omega_p }{(2\pi)^3} \int d^3 \vec{r}_1\; d^3
  \vec{r}_1^{\,\prime}\; d^{3(A-1)} \{\vec{r}_{2-A}\}
  e^{-i\vecp\cdot(\vec{r}^{\,\prime}_1 - \vec{r}_1)}
  \Psi^*(\vec{r}_1,\vec{r}_{2-A})
  \Psi(\vec{r}_1^{\,\prime},\vec{r}_{2-A}).  
\end{equation}
The corresponding
single-nucleon operator $\hat{n}_{p}$ reads
\begin{equation}
 \hat{n}_p =
  \frac{1}{A} \sum_{i=1}^A \int \frac{d^2 \Omega_p }{(2\pi)^3}
  e^{-i\vecp\cdot(\vec{r}^{\,\prime}_i - \vec{r}_i) } = 
\sum_{i=1}^A
  \hat{n}^{[1]}_{p} \left( \vec{r}_i , \vec{r}^{\, \prime}_i \right) 
=
\sum_{i=1}^A
  \hat{n}^{[1]}_{p}(i) \; .
\end{equation}
 The operator $ \hat{n}_p $ and the expansion of \eqn{omeff2}
 determine an effective two-body operator $\hat{n}_p^{\text{LCA}}$
 from which the correlated single-nucleon momentum distributions at
 momentum $p$ can be computed.  The operator $\hat{n}_p^{\text{LCA}}$
 can be evaluated in the IPM ground-state wave function.  The diagrams
 in figure~\ref{fig:diagrams} are a schematic graphical
 representation of the different contributions to $n^{[1]}(p)$ after
 introducing the effective operator $\hat{n}_p^{\text{LCA}}$.

In order to preserve the normalisation properties $\int dp \; p^{2}
n^{[1]}(p) =1$ in the LCA, the normalisation factor $\mathcal{N}$ of
\eqn{defpsi} is expanded up to the same order as the operator of
\eqn{omeff2}, 
\eqa[normexp]{ \mathcal{N}  = 1 + \frac{2}{A}
  \sum_{\alpha<\beta}  {}_{\text{nas}} \bra{\alpha\beta}
  \hat{l}^\dagger(1,2)+ \hat{l}^\dagger(1,2)\hat{l}(1,2) +
  \hat{l}(1,2) \ket{\alpha\beta}_{\text{nas}} .}  
Here, $\ket{\alpha\beta}_{\text{nas}}$ is the uncoupled normalised and
anti-symmetrized (nas) two-nucleon state in the $\left( \vec{r}_1, \vec{r}_2 \right)$-space.  The summation
$\sum\limits_{\alpha<\beta}$ extends over all occupied single-nucleon
states. Those states are identified by the quantum numbers $\alpha
\equiv n_\alpha l_\alpha j_\alpha m_{j_\alpha} t_\alpha$, whereby
$t_\alpha$ denotes the isospin projection.

\begin{table}[tb]
\caption{The norm $\mathcal{N}$ of \eqn{normexp} for a wide range of nuclei. \label{tab:norms}}
\begin{indented}
\item[]\begin{tabular}{@{}llll}
\br
    $^{2}$H &   $1.128$ & 
    $^{40}$Ca & $1.637$ \\ 
    $^{4}$He &  $1.327$ & 
    $^{48}$Ca & $1.629$ \\ 
    $^{9}$Be &  $1.384$ & 
    $^{56}$Fe & $1.638$ \\ 
    $^{12}$C &  $1.435$ & 
    $^{108}$Ag & $1.704$ \\ 
    $^{16}$O &  $1.527$ & 
    $^{197}$Au & $1.745$ \\
    $^{27}$Al & $1.545$ &  
    $^{208}$Pb & $1.741$ \\ 
\br
  \end{tabular}
\end{indented}
\end{table}

In order to construct the IPM single-particle wave functions we adopt
a harmonic oscillator (HO) basis with a global mass-dependent parametrisation 
\begin{equation}
\hbar\omega = 45A^{-1/3} - 25
A^{-2/3}.  
\label{eq:paramho}
\end{equation}
In a HO basis, a transformation from
$(\vec{r}_1,\vec{r}_2)$ to $(\vec{r}_{12}=\vec{r}_1 - \vec{r}_2
,\vec{R}_{12} = \frac{\vec{r}_1 + \vec{r}_2}{2})$ for the nas
two-nucleon state can be readily performed
\cite{Vanhalst:2012ur,Vanhalst:2011es}
\begin{equation}
  \ket{ \alpha  \beta}_{\text{nas}}  
  =
  \sum_{D} \ket{D} \braket{D}{\alpha\beta}_{\text{nas}}
   ,
  \label{eq:transformation}
\end{equation}
where we have introduced a shorthand notation for the quantum
states of the pairs in the $\left( \vec{r}_{12}, \vec{R}_{12} \right) $ coordinate space 
\begin{equation}
 \ket{D} \equiv \ket{nlSjm_j, NLM_L , T M_T} \; .
\label{eq:comcm}
\end{equation}
Here, $n$ and $l$ are the radial
and orbital angular-momentum quantum numbers corresponding with the
relative motion of the pair. The $jm_j$ are the quantum numbers of the
total angular momentum of the pair. The $TM_T$ $(S)$ determine the
isospin (spin) quantum numbers of the pair.  The c.m.~wave function is
described by the quantum numbers $NLM_L$.

\begin{figure}
  \begin{center}
  \includegraphics[width=0.60\textwidth]{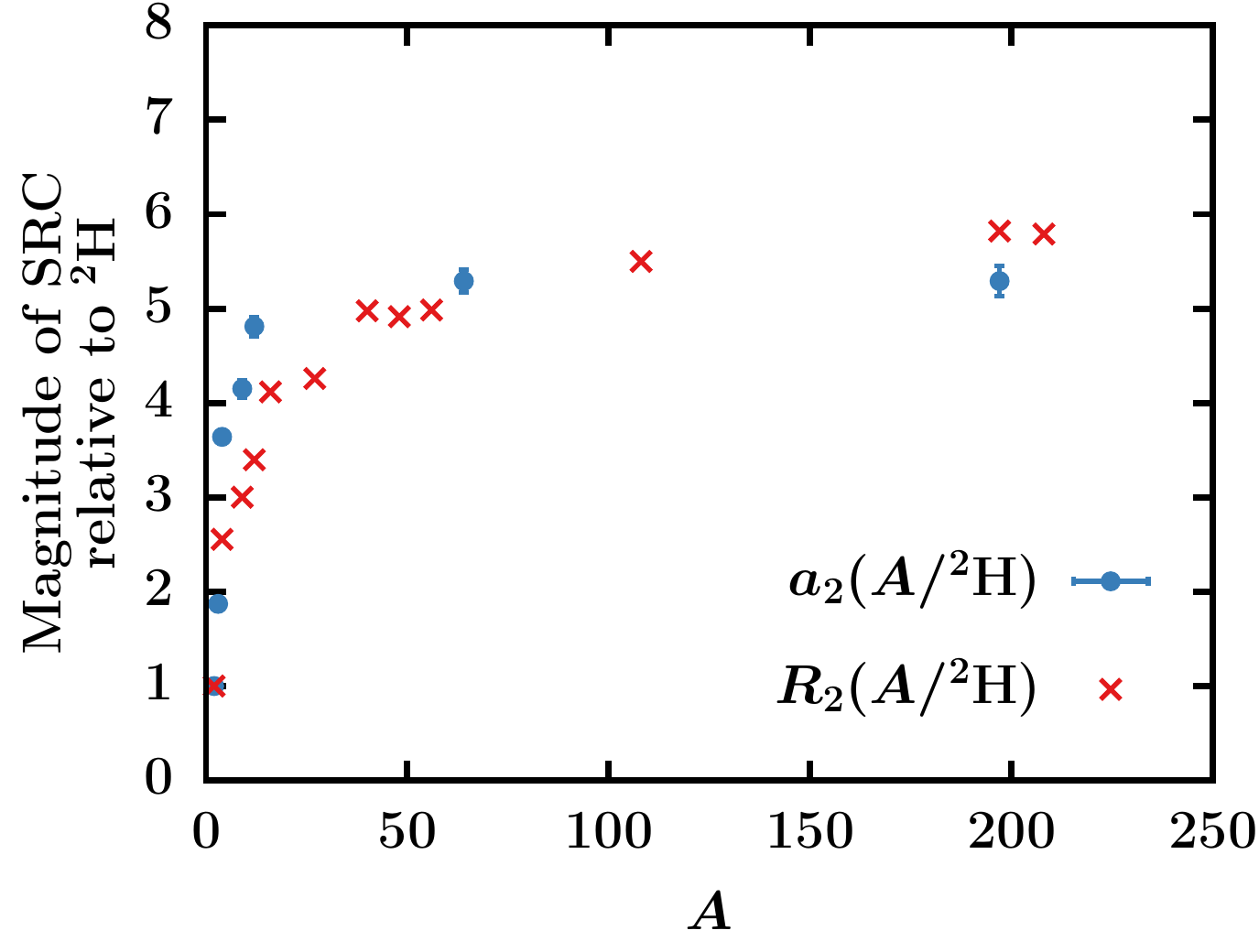}
  \end{center}
  \caption{ The mass dependence of the computed ratios
    $R_{2}(A/{}^2\text{H})$ defined in equation~(\ref{eq:R2}) and of the
    experimentally extracted $a_2(A/{}^2\text{H})$ coefficients from
    \cite{PhysRevC.85.047301}.
    \label{fig:a2norm}
  }
 \end{figure}

\begin{figure}
  \begin{center}
  \includegraphics[width=0.60\textwidth]{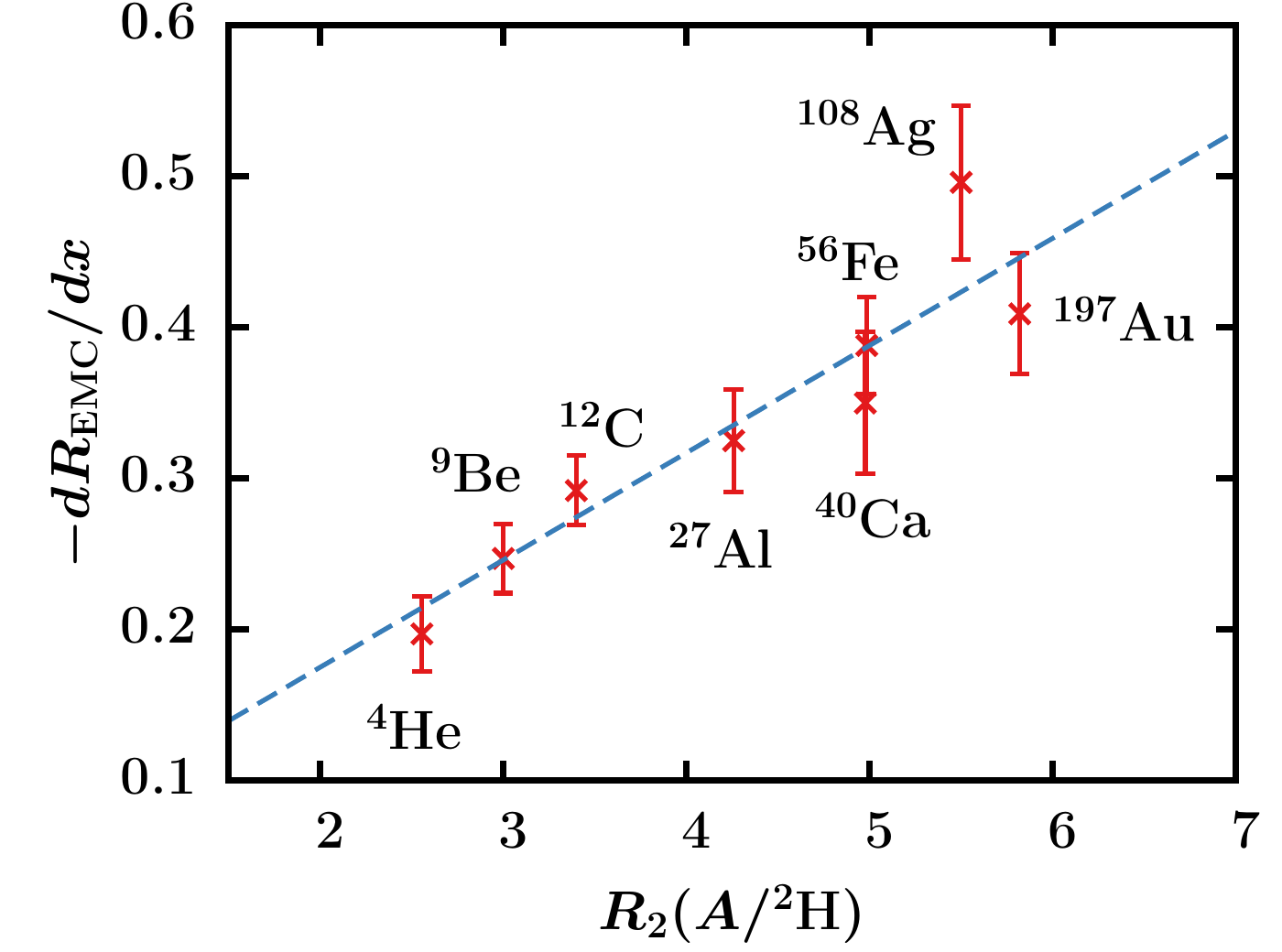}
\end{center}  
\caption{The measured magnitude of the EMC effect, $- \frac{ d
      R_{EMC} }{ d x}$ is plotted as a function of the computed
    $R_{2}(A/{}^2\text{H})$ ratios defined in equation~(\ref{eq:R2}).
    The values of the EMC magnitude are from the analysis presented in
    \cite{PhysRevC.86.065204}.  The fitted dashed line obeys the
    equation $- \frac{ d R_{EMC} }{ d x} = (0.033 \pm 0.035) + ( 0.071
    \pm 0.009 ) \cdot R_2(A/{}^2\text{H}) $. 
    \label{fig:EMC}
  }
 \end{figure}

\begin{figure}
  \begin{center}
  \includegraphics[angle=0,width=\textwidth]{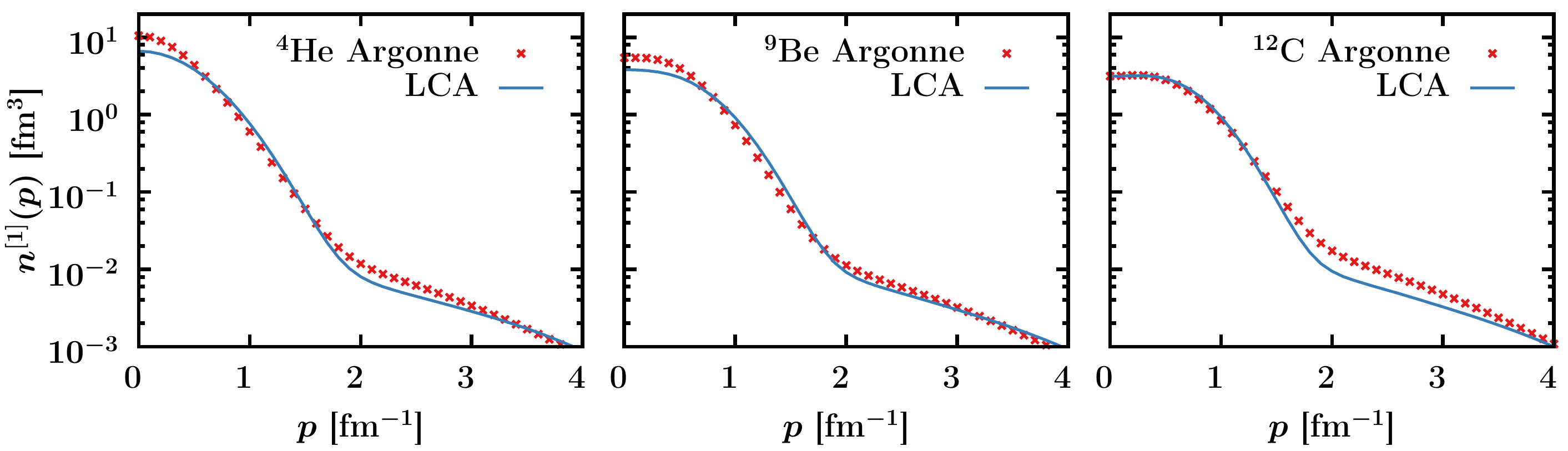}
  \end{center}
  \caption{ The momentum dependence of the $n^{[1]}(p)$
    for $^{4}$He, $^{9}$Be and $^{12}$C.  The red 
  crosses  are the QMC results of  
  \cite{PhysRevC.89.024305} obtained with the Argonne $v_{18}$ two-nucleon and Urbana X three-nucleon potentials.
    \label{fig:compAr}
  }
 \end{figure}

Table~\ref{tab:norms} lists the computed values of the normalisation
factors of \eqn{normexp} for a range of nuclei from $^{2}$H to
$^{208}$Pb. The model dependence of the computed $\mathcal{N}$ is
related to the choices made with regard to the IPM basis and the
correlation functions.  Tests for a few nuclei indicate that replacing
the HO basis by a Woods-Saxon one changes the computed $\mathcal{N}$
by a few percent. This is connected with the observation that the
amount of close-proximity nucleon pairs in a nucleus is rather
insensitive to the choice of the single-particle wave functions
\cite{Vanhalst:2012ur}. The sensitivity of the computed $\mathcal{N}$
to the choices with regard to the correlation functions is larger. For
example, after switching off the effect of spin-isospin correlations
we find a $\mathcal{N}$ which is about 5\% smaller for $^2$H and about 10\%
smaller for the medium-heavy and heavy nuclei listed in
table~\ref{tab:norms}.

The deviation of $\mathcal{N}$ from 1 can be
interpreted as a quantitative measure for the total effect of the SRC
operators on the IPM ground-state wave function.  For the deuteron,
the tensor correlation operator acting on the relative S-wave of the
IPM nucleon pair wave function is responsible for the D-wave
component.  The LCA is a crude approximation for the proton-neutron
deuteron system. Nevertheless, the tail part of the LCA deuteron
momentum one-body momentum distribution is in fair agreement with the
realistic WCJ1 one \cite{Gross:2010qm}, which has a 7.3\% D-wave
admixture.
%
The $a_2(A/{}^2\text{H})$ coefficient is an experimentally determined
quantity which is connected with the magnitude of SRC in nucleus $A$
relative to $^{2}$H \cite{PhysRevLett.108.092502,PhysRevC.48.2451}.
It is extracted from the scaling behaviour of the measured
$A(e,e^{\prime})/^{2}\text{H}(e,e^{\prime})$ cross-section ratio in
selected kinematics favouring virtual-photon scattering from correlated
pairs.  In figure~\ref{fig:a2norm}, the ratios of the computed norms for
$A$ relative to $^{2}$H
\begin{equation}
  R_2(A/{}^2\text{H}) =  \frac{ \mathcal{N}(A) -1}{ \mathcal{N}({}^2\text{H}) - 1} \; ,
  \label{eq:R2}
\end{equation}
are compared to the measured $a_2$ coefficients of \cite{PhysRevC.85.047301}.  In the framework developed in this
work, the $R_2(A/{}^2\text{H})$ are a measure of the magnitude of the
aggregated effect of SRC in nucleus $A$ relative to their magnitude in
$^2\text{H}$.  As can be appreciated from figure~\ref{fig:a2norm}, the
mass dependence of the measured $a_2$ and computed
$R_2(A/{}^2\text{H})$ ratios is roughly the same.  For $A \lesssim
40$, $R_2(A/{}^2\text{H})$ increases strongly with mass number $A$
which hints at a strong mass dependence of the quantitative effect of
SRC.  For $A>40$, the predicted mass dependence of the magnitude of
the SRC is soft.

Recently, it has been suggested that the magnitude of the European
Muon Collaboration (EMC) effect in a specific nucleus $A$ is connected
with the magnitude of the SRC in $A$~\cite{PhysRevLett.106.052301}.
Consequently, one can expect a linear relation between the $R_2$ of
equation~(\ref{eq:R2}) and the magnitude $ - \frac{ d R_{EMC} }{ d x}
$ of the EMC effect.  This suggestion is clearly confirmed in
figure~\ref{fig:EMC} which illustrates the correlation between the
experimentally extracted $ - \frac{ d R_{EMC} }{ d x} $ and the LCA
predictions for the aggregated effect of SRC in nucleus $A$ relative
to $^2$H. Clearly, the observed correlation does not imply causation.

\section{Single-nucleon momentum distribution}
\label{sec:obmd}
In figure~\ref{fig:compAr} we compare the LCA results for the
$n^{[1]}(p)$ with those obtained with quantum Monte-Carlo (QMC)
methods using realistic two-nucleon and three-nucleon Hamiltonians
\cite{PhysRevC.89.024305}.  With the normalization factor of
(\ref{eq:normexp}), the single-nucleon momentum distributions are
normalized as $1= \int dp \; p^{2} n^{[1]}(p)$, which facilitates the
comparison over the various nuclei.
Up to the characteristic nuclear Fermi momentum $p_F=1.25$~\invfm, the
shape of $n^{[1]}(p)$ is very Gaussian in both approaches. For $p>p_F$
the distributions are heavy-tailed. For $p \gtrsim 3$~\invfm, the QMC
and the LCA method predict a comparable exponential-like fat tail,
which is very remarkable given the very different frameworks in which
the results are obtained. For medium momenta $p \approx 2$~\invfm the
LCA predictions for the $n^{[1]}(p)$ undershoot the QMC ones. This can
be attributed to the lack of LRC in the LCA framework. Indeed,
the effect of LRC is known to extend to medium
nucleon momenta \cite{Feldmeier:2011tt,Dickhoff:2004xx}. In the same vein, it is not surprising that for
$^4$He and $^9$Be the LCA and QMC display some differences at low $p$,
given that LCA does not account for the complicated long-range cluster
structures of those nuclei. 
In this context, it is worth mentioning that
the nuclear-matter studies of \cite{Rios:2013zqa} have clearly
illustrated that the fat tails of the single-nucleon distributions are
sensitive to the adopted realistic nucleon-nucleon interaction.  This
is related to the fact that the short-range part of the $NN$ force is
not well constrained by a fit to $NN$ scattering data.

\begin{figure}
\begin{center}
  \includegraphics[angle=0,width=\textwidth]{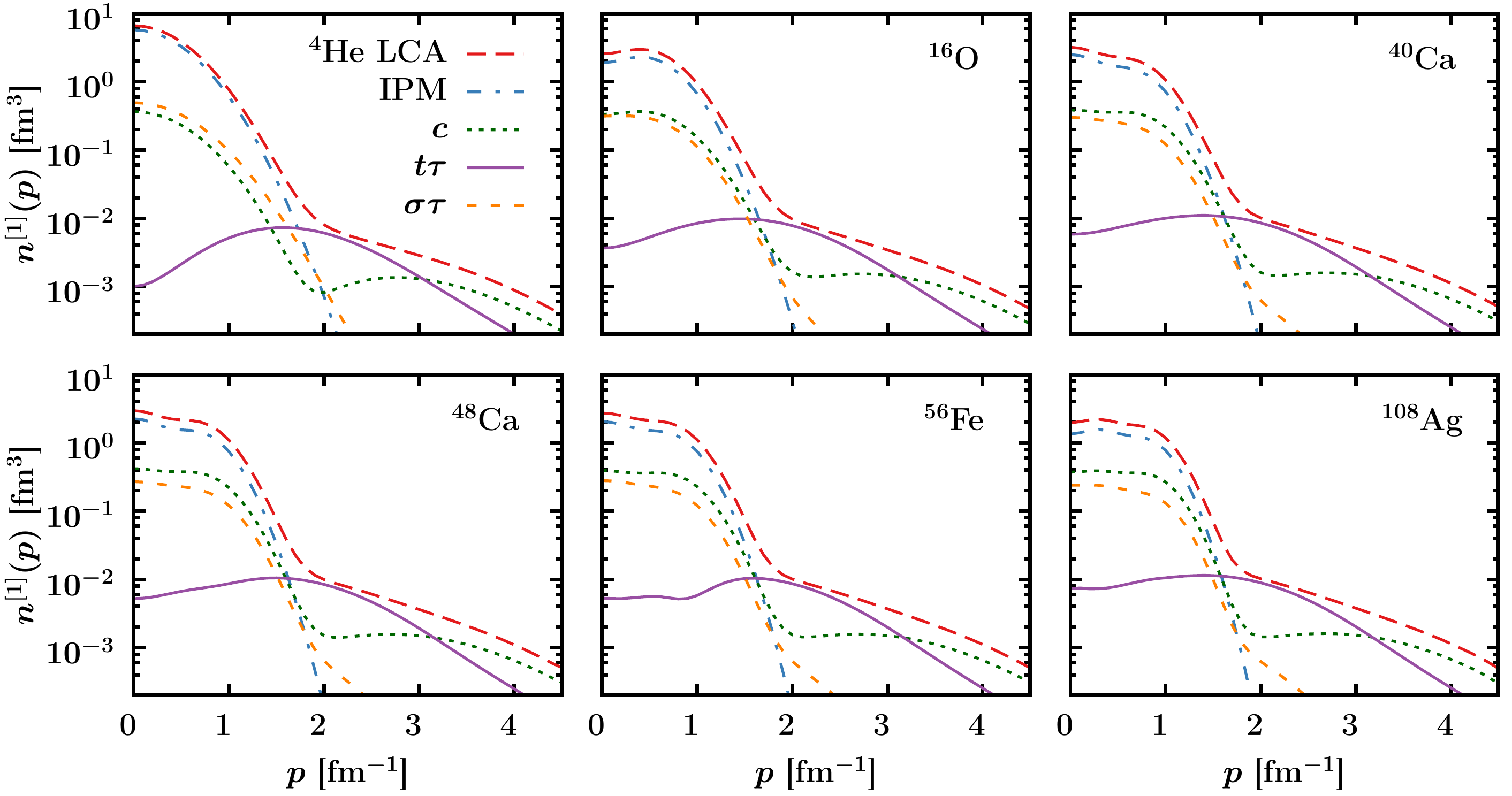}
\end{center}
   \caption{\label{fig:momdiscorr}  The single-nucleon
     momentum distribution $n^{[1]}(p)$ for six nuclei. The long
     dashed line is the full LCA result. The dashed-dotted line is the
     IPM contribution to the LCA result. Also shown are the results of
     a calculation that only includes the two-body central (green
     dotted line), tensor (purple solid line) and spin-isospin (orange
     short-dashed line) correlation contribution.  The LCA result
     includes the interference between all contributions.}
\end{figure}

The LCA results for the $n^{[1]}(p)$ are displayed in
figure~\ref{fig:momdiscorr} for a range of nuclei from He to Ag. Some
stylized features which apply to all studied nuclei are emerging from
the LCA calculations. For $p \lesssim 1.5$~fm$^{-1}$ the distribution
is dominated by the IPM contribution (diagram (a) of
figure~\ref{fig:diagrams}) and the SRC do not affect the momentum
dependence of $n^{[1]}(p)$. The fat tails are induced by the
correlations (diagrams (b), (c) and (d) of figure~\ref{fig:diagrams})
whereby one distinguishes two regions. For $1.5 \lesssim p \lesssim
3$~\invfm the tensor correlations dominate. The effect of the central
correlations extends over a large momentum range and for $p>
3.5$~\invfm, it represents the dominant contribution to $n^{[1]}(p)$
(with the tensor part gradually losing in importance).  For all nuclei
the crossover between the tensor and the central correlated part of
the tail of $n^{[1]}(p)$ occurs at a momentum slightly larger than
3~fm$^{-1}$. At momenta approaching 4~fm$^{-1}$ the central
correlations provide about half of the the $n^{[1]}(p)$ while the
remaining strength is almost exclusively due to the interference
between the central and spin-isospin correlations (not shown
separately in figure~\ref{fig:momdiscorr}). This qualitative behaviour
is in line with the ab-initio $^4$He results of
\cite{PhysRevC.89.024305} (see figure~3 of that reference).  The
above-mentioned conclusions which apply to the correlated part of the
one-body momentum distributions of all nuclei studied here, are
qualitatively in line with the nuclear-matter results of
\cite{Rios:2013zqa}. This illustrates that the effect of SRC on
single-nucleon momentum distributions can be summarised in some
universally applicable principles.

\begin{figure}
  \begin{center}
  \includegraphics[angle=0,width=\textwidth]{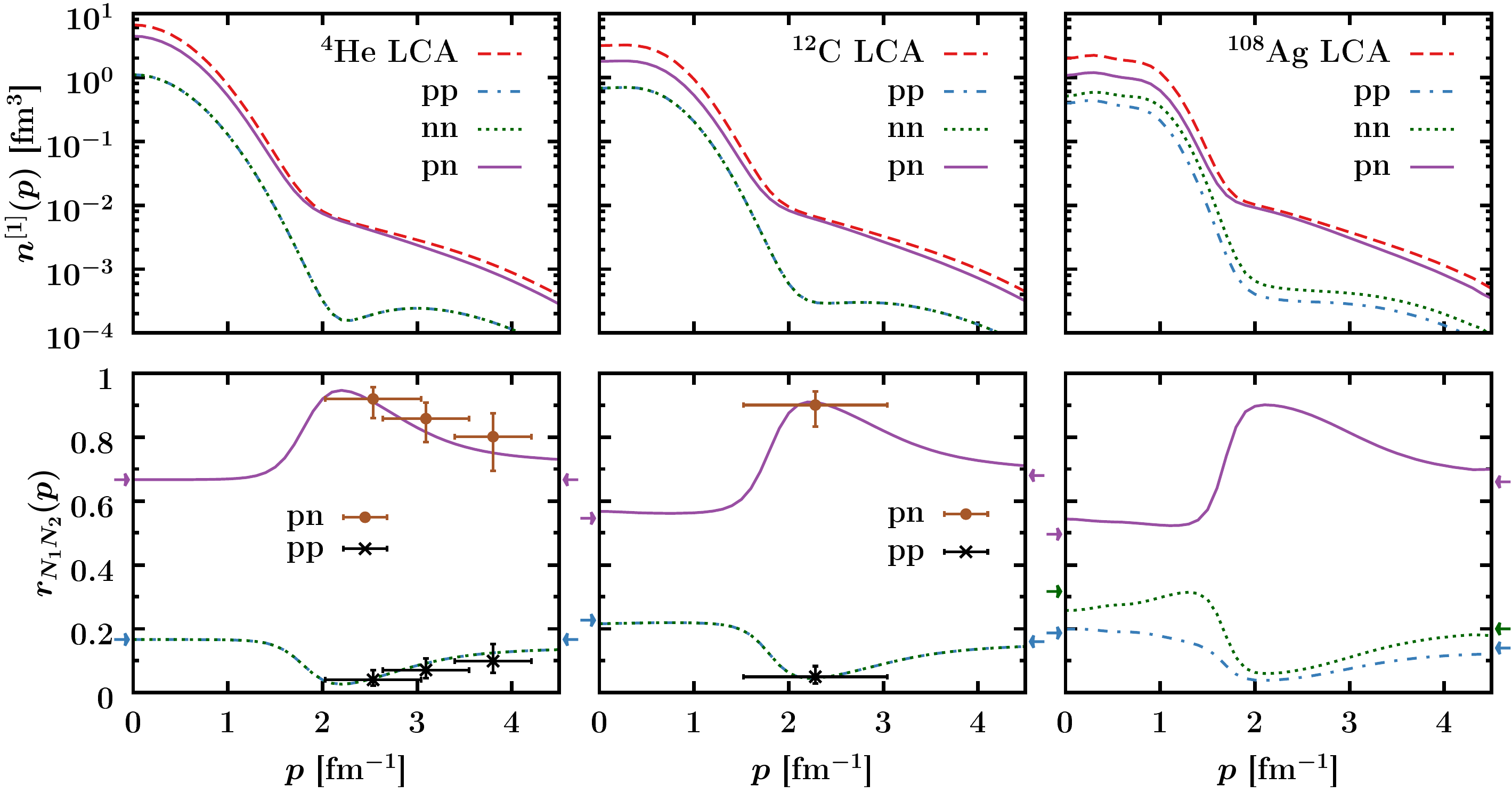}
  \end{center}
  \caption{ The top panels show the LCA results for the
    momentum dependence of the contribution of $pp$ pairs
    ($n^{[1]}_{pp}(p)$), $nn$ pairs ($n^{[1]}_{nn}(p)$), and $pn$
    pairs ($n^{[1]}_{pn}(p)$) to the $n^{[1]}(p)$ of $^{4}$He,
    $^{12}$C and $^{108}$Ag. The bottom panels show the momentum
    dependence of the ratios $r_{N_1 N_2} = n^{[1]}_{N_1N_2}(p)/
    n^{[1]}(p)$ for $N_1N_2=pp,nn,pn$.  The data points for $^4$He are
    extracted from the bottom panel of figure~$2$ in
    \cite{Korover:2014dma}.  The data points for $^{12}$C are
    extracted from \cite{Subedi:2008zz}. For $^{4}$He and
    $^{12}$C the theoretical results for $pp$ overlap almost perfectly
    with those for $nn$. The arrows at $p=0$ indicate the naive IPM predictions for the $r_{N_{1}N_{2}}$. The arrows at $p \approx 4.5$~\invfm are the predictions for the $r_{N_{1}N_{2}}$ based on the counting of the $nl=00$ pairs (see text for more details).  
    \label{fig:momdisxx}
   } 
 \end{figure}

The dominant role of the tensor correlations for intermediate nucleon
momenta $1.5 \lesssim p \lesssim 3$~\invfm, has some important
implications for the isospin dependence of the effect of short-range
correlations. With the aid of \eqn{omeff2} one can write
\begin{equation}
n^{[1]}(p)=n^{[1]}_{pp}(p)+n^{[1]}_{nn}(p)+n^{[1]}_{pn}(p) ,
\end{equation}
with 
\begin{equation}
 n^{[1]}_{N_1N_2}(p) = \frac{1}{\mathcal{N}}
  \sum_{\alpha<\beta} \delta_{t_\alpha, N_1} \delta_{t_\beta, N_2} 
  {}_{\text{nas}} \bra{\alpha\beta} \hat{n}^{\text{LCA}}_p(1,2)
  \ket{\alpha\beta}_{\text{nas}}.   
\label{eq:NNmom}
\end{equation}
Referring to figure~\ref{fig:diagrams}, the $n^{[1]}_{N_1N_2}(p)$
encodes how much the $pp$, $nn$ and $np$ pairs $\left( \alpha, \beta
\right)$ contribute to $n^{[1]}$ at given $p$.  The LCA results for
$n^{[1]}_{N_{1}N_{2}}(p)$ are shown in figure~\ref{fig:momdisxx}.
The ratio $r_{N_{1}N_{2}} (p) \equiv
n^{[1]}_{N_{1}N_{2}}(p)/n^{[1]}(p)$ quantifies the relative
contribution of $N_1N_2$ pairs to $n^{[1]}(p)$ at given momentum
$p$. In a naive IPM one expects momentum-independent values of
$r_{pp}=\frac{Z(Z-1)}{A(A-1)}$, $r_{nn}=\frac{N(N-1)}{A(A-1)}$ and
$r_{pn}=\frac{2NZ}{A(A-1)}$. For $p<p_{F}$ the plotted ratios in the
bottom panel of figure~\ref{fig:momdisxx} very much follow these naive
expectations.  The tensor dominated momentum range is characterised by
an increase of the $pn$ contribution to $n^{[1]}(p)$.

The above discussion provides a natural explanation for the
observation that SRC-sensitive reactions like two-nucleon knockout
($A(e,e'pN)$ and $A(p,ppN)$ reactions for example) are very much
dominated by the $pn$ channel in the tensor-dominated region which
roughly corresponds with $1.5 \lesssim p \lesssim 3$~\invfm.  The
bottom panels of figure~\ref{fig:momdisxx} suggest that under those
conditions the $pn$ channel can represent 90\% of the correlated
strength, leaving a mere 5\% for the $pp$ channel. This prediction
seems to be in line with the experimental observations.

Indeed, the small ratio of $pp$-to-$np$ pairs above the Fermi momentum
has been recently established in $^{12}$C$(e,e'p(p))$,
$^{27}$Al$(e,e'p(p))$, $^{56}$Fe$(e,e'p(p))$ and
$^{208}$Pb$(e,e'p(p))$, measurements at Jefferson Lab
\cite{Hen:2014,Subedi:2008zz}.  The quoted $pp$ to $pn$ ratio of
$\frac { 1 \pm 0.3 \%} {18 \pm 5\%}$ for $^{12}$C, displayed in
figure~\ref{fig:momdisxx} is compatible with the LCA predictions
thereby assuming that the $pp$ and $nn$ contributions are equal for
$N=Z$ nuclei.  From an analysis of the ratio
$\frac{^{12}\text{C}(p,ppn)} {^{12}\text{C}(p,pp)}$ it could be
inferred that the removal of a proton from the nucleus with initial
momentum 275--550 MeV/c is 92$^{+8}_{-18}$\% of the time accompanied
by a neutron \cite{Piasetzky:2006ai}.  Also this result is in line
with the LCA predictions for $^{12}$C contained in
figure~\ref{fig:momdisxx}.  Our results indicate that similar
anomalously large $r_{pn}/r_{pp}$ ratios may be found for heavier
nuclei when probing the tensor-dominated tail of the single-nucleon
momentum distribution.

Another interesting feature of the results of figure~\ref{fig:momdisxx}
is that the $r_{pp}(p)$ [$r_{pn}(p)$] reaches its minimum (maximum) at
$p\approx 2$~fm$^{-1}$. For $p >2$~fm$^{-1}$ the $r_{pp}(p)$ grows and
the $r_{pn}(p)$ decreases. Experimental evidence supporting this
prediction has been recently obtained in the simultaneous measurement
of exclusive $^{4}$He$(e,e'pp)$ and $^{4}$He$(e,e'pn)$ at $(e,e'p)$
missing momenta from 2 to 4.3~\invfm \cite{Korover:2014dma}. In those
measurements, the kinematics is tuned to probe a nucleon at a given
momentum $p>p_F$ in conjunction with its correlated partner. These are
precisely the SRC induced two-nucleon processes which systematically
dominate the LCA $n^{[1]}(p)$ above the Fermi momentum. One may be
tempted to connect A$(e,e'pN)$ cross sections to two-nucleon momentum
distributions (TNMD). First, even after cross-section factorisation no
direct connection between the cross sections and TNMD can be
established \cite{PhysRevC.89.024603}. Second, as has been pointed out
in \cite{Feldmeier:2011tt} (a nice pictorial description is given
in Figure~12 of that reference) the correlated part of the TNMD
receives large SRC contributions from three-nucleon
configurations. Thereby the correlation is mediated through a third
nucleon. The exclusive A$(e,e'pN)$ measurements are not kinematically
optimised to probe those three-nucleon configurations. The A$(e,e'pN)$
kinematic settings are optimised to probe SRC-related two-nucleon
configurations, and it is precisely those configurations which are the
source of strength of the tails of the single-nucleon momentum
distributions.

The $^4$He data points shown in figure~\ref{fig:momdisxx} are extracted
from the $^{4}$He$(e,e'pp)$/$^{4}$He$(e,e'pn)$ cross-section ratios of
\cite{Korover:2014dma}, whereby we have assumed that
$r_{nn}=r_{pp}$.  The $r_{np}$ and $r_{pp}$ cannot be directly
connected to the $^{4}$He$(e,e'pp)$/$^{4}$He$(e,e'p)$ and
$^{4}$He$(e,e'pn)$/$^{4}$He$(e,e'p)$ cross-section ratios also shown in
figure~2 of~\cite{Korover:2014dma}. Indeed, for $p>p_{F}$ the
$r_{N_{1}N_{2}}(p)$ encodes information about correlated pairs,
whereas the $^{4}$He$(e,e'p)$ cross sections also contain
contributions from other sources like final-state interactions and
triple correlations.

For $p>p_F$ the
  ratio $\frac {r_{pn}(p)}{{r_{pn}(p)+r_{pp}(p)}}$ can be interpreted
  as the ratio of the number of SRC proton-neutron pairs to the sum of
  the proton-neutron and proton-proton ones at a momentum $p$. In \cite{Hen:2014} this ratio has been extracted from the combination
  of $A(e,e'pp)$ and $A(e,e'p)$ measurements for the nuclei $^{12}$C,
  $^{27}$Al, $^{56}$Fe and $^{208}$Pb.  The experimental values for the ratios are $\approx
  0.95$ for the four nuclei and are extracted over a bin covering the
  range $1.5 \le p \le 4.5$~fm$^{-1}$. Our calculations, reproduce the
  observations that for $p> p_F$ the $\frac
  {r_{pn}(p)}{{r_{pn}(p)+r_{pp}(p)}}$ are rather mass independent and
  adopt a value indicative of the dominance
  of the proton-neutron SRC pairs.

As the central correlations, which are blind
for the isospin of the interacting pairs, gain in importance with
increasing $p$ one observes in figure~\ref{fig:momdisxx} that the $r_{N_{1}N_{2}} (p)$ ratios
gradually approach a limiting value which is different from the IPM
values, in particular for heavier nuclei. 

The above discussions indicate that the LCA framework in combination
with central, tensor and spin-isospin correlations, captures the
stylized features of the SRC including its mass and isospin
dependence.  We now wish to shed light on the underlying physics
mechanics of the correlated part of the momentum distribution. More in
particular we address the question: ``What are the quantum numbers of
the IPM pairs which are most affected by the correlations?'' This
discussion will lead to an understanding of the high $p$ limits in the
bottom panels of figure~\ref{fig:momdisxx}.

One can determine the contributions from the relative quantum numbers
$n l$ of the IPM pairs to the correlated part of $n^{[1]}(p)$ 
(denoted by ${n}^{[1],\text{corr}}(p)$)  by means of the
expansion of~(\ref{eq:transformation}). One finds,
\begin{equation}
\fl
 {n}^{[1],\text{corr}}_{n l, n^{\prime} l ^{\prime}}(p) = {}
  \sum_{\alpha< \beta} \sum_{D,E}
\left[ \braket{D}{\alpha\beta}_{\text{nas}} \right] ^ {\dagger}
\braket{E}{\alpha\beta}_{\text{nas}}
  \delta_{n n_D} \delta_{l l_D} \delta_{n^{\prime} n_E}
  \delta_{l^{\prime} l_E}  \bra{D}
  \hat{n}^{[1],\text{corr}}_{p}(1,2) \ket{E} \; , 
\end{equation} 
where the operator $\hat{n}^{[1],\text{corr}}_{p}(1,2)$ and the states $\ket{E}, \ket{D}$ have been
defined as in~(\ref{eq:correlatedpartofomega}) and (\ref{eq:comcm}).  Obviously, one
has 
\eq{ \sum_{nl} \sum_{n^{\prime} l ^{\prime}}
  n^{[1],\text{corr}}_{n l , n^{\prime} l ^{\prime}}
  (p) = n^{[1],\text{corr}}(p) \; .
}

\begin{figure}
\begin{center}
  \includegraphics[angle=0,width=\textwidth]{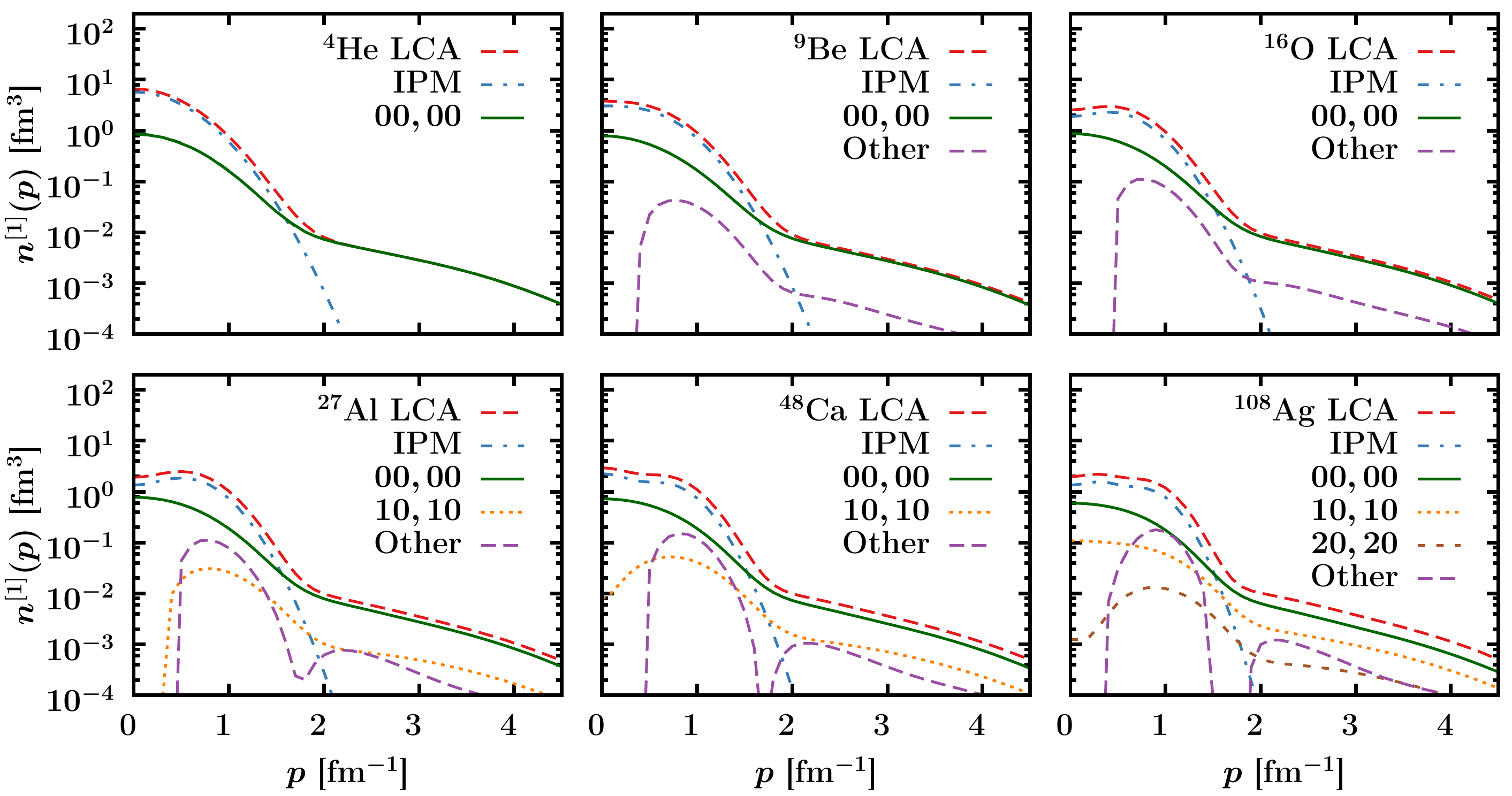}
\end{center}
  \caption{ The momentum dependence of the $n^{[1]}(p)$
    for six nuclei. The red dashed line is the LCA result. The blue
    dashed-dotted line is the IPM contribution to the LCA result. Also
    shown are the $n^{[1],\text{corr}}_{nl,n^{\prime} l ^{\prime}}(p)$
    that dominate the high momentum tail.  The purple dashed line is
    the summed contribution of the the
    $n^{[1],\text{corr}}_{nl,n^{\prime} l ^{\prime}}(p)$ which are not
    shown separately.
  \label{fig:momdis} } 
 \end{figure}

The $n^{[1],\text{corr}}_{nl,n^{\prime} l^{\prime}}(p)$ that provide
the largest contribution to $n^{[1]}(p)$ are shown in
figure~\ref{fig:momdis}.  It is clear that correlation operators acting
on $nl=00$ IPM pairs are responsible for the major fraction of the
$n^{[1]}(p)$ for $p \gtrsim \;$2~fm$^{-1}$.  For heavier nuclei, the
contributions from pairs with $n>0$ and $l=0$ gain in importance.
Non-diagonal $\hat{n}^{[1],\text{corr}}_{nl,n^{\prime} l
  ^{\prime}}(p)$ represent a small fraction of the high-momentum tail.

We wish to stress that correlation operators acting on IPM pairs can
change the quantum numbers. For example, the tensor operator acting on
the deuteron's $l=0$ IPM pair generates the correlated $l=2$ state.
The dominant role of $nl=00$ IPM pairs in the creation of
high-momentum components, provides support for our proposed method to
quantify the SRC by counting the number of $nl=00$ IPM pairs
~\cite{Vanhalst:2012ur,Vanhalst:2011es,PhysRevC.89.024603}.
Consequently, for high $p$ the central correlations dominate and the
$r_{N_1 N_2}(p)$ ratios of figure~\ref{fig:momdisxx} are connected
with the amount of $N_1N_2$ IPM pairs with $nl=00$.  Using the
computed number of of $nl=00$ pairs in $^{12}$C we find $r_{pp} =
r_{nn} = 0.16$ and $r_{pn}= 0.68$. For $^{108}$Ag, a similar
calculation leads to $r_{pp}=0.14$, $r_{nn}= 0.20$ and $r_{pn}=0.66$.
For high $p$ these numbers are fair predictions for the computed
ratios $r_{N_1 N_2}(p)$ in figure~\ref{fig:momdisxx}. The dominant
role of the $nl=00$ pairs in generating the high-momentum components
of the single-nucleon momentum distributions provides also a natural
explanation for the observation that the high-momentum tail of the
single-nucleon momentum distributions of nuclei has a universal
shape. Indeed, the wave function for the $nl=00$ pairs does not
dramatically change as one moves through the nuclear mass table.

\section{Single-nucleon kinetic energies and rms radii}
\label{sec:kinene}
We now turn to a discussion of the LCA predictions for the
single-nucleon kinetic energies $\left< T_{N} \right>$ and rms
radii. The $\left< T_{N} \right>$ are not observables but the LCA
results can be compared with previously published ones. In addition, in recent
publications \cite{OrTensor,Carbone}, the isospin dependence of the
$\left< T_{N} \right>$ has been connected with the kinetic part of of
the nuclear symmetry energy, which is one of the key bulk properties
of atomic nuclei.  In a non-relativistic framework, the diagonal
single-nucleon kinetic energy operator $\widehat{T}^{[1]}$ can be
written as \eq{ \widehat{T}^{[1]} = \sum_{i=1}^A \widehat{T}^{[1]} (i)
  = \sum_{i=1}^A \frac{- \hbar ^{2}}{2 M_{i}} \nabla^2_i, } where
$M_{i}$ is the nucleon mass. In the IPM, the average kinetic energy $
\left< T_{p} \right>$ per proton is given by
\eq[kinIPM]{ \left<
  T_{p}^{IPM} \right> = \frac{1}{Z} \sum_{\alpha} \delta_{t_\alpha,p}
  \bra{\alpha} \widehat{T}^{[1]}(1) \ket{\alpha} \; .  
}%
A similar
definition is adopted for the average kinetic energy per neutron
$\left< T_{n} \right>$. In the LCA framework developed in
section~\ref{sec:formalism} one has \eq[kinLCA]{ \left<
  T_{p}^{\text{LCA}} \right> = \frac{1}{\mathcal{N}} \frac{1}{Z}
  \sum_{\alpha<\beta} {}_{\text{nas}} \bra{\alpha\beta}
  \widehat{T}^{\text{LCA}}_p(1,2) \ket{\alpha\beta}_{\text{nas}}, }
where the operator $\widehat{T}_{p}^{\text{LCA}}$ can be obtained from
\eqn{omeff2}.  Since we work in a non-relativistic framework, we
have adopted a hard cutoff of $4.5$~\invfm for the maximum nucleon
momentum in the calculations of the kinetic energy.
\begin{table}[tb]
  \caption{Results from the IPM and LCA framework for the kinetic
    energy per proton and neutron ($\left< T_{p} \right>$ and $\left<
    T_{n} \right>$) for a variety of nuclei. We compare to values
    obtained for the average correlated kinetic energy per nucleon $\left< T_{N}
    \right>$ from alternate calculations
    \cite{Claudio1996,Feldmeier:1997zh}.
  \label{tab:kinenergy}
}

\begin{indented}
\item[]\begin{tabular}{@{}cccccccccc}
\br
   &  & \multicolumn{6}{c}{$\left< T_N \right>$~(MeV)} & 
\multicolumn{2}{c}{$\left< T_p \right>/\left< T_n \right>$} \\
& & \crule{6} & \crule{2} \\
  $A$ & $x_p = \frac{Z} {A}$ & IPM (p) & IPM (n) & LCA (p) & LCA(n) & \cite{Claudio1996} & \cite{Feldmeier:1997zh}  & IPM & LCA \\
\mr   
    $^{2}$H &   
    $0.500$ &
    $14.95$ &
    $14.93$ & 
    $20.95$ &
    $20.91$ &
          &
          &
    $1.00$ &
    $1.00$ \\
    $^{4}$He &  
    $0.500$ &
    $13.80$ &
    $13.78$ &
    $25.28$ &
    $25.23$ &
           &
    $19.63$ &
    $1.00$ &
    $1.00$ \\
    $^{9}$Be &
    $0.444$ &
    $15.81$ &
    $16.58$ &
    $28.91$ &
    $27.33$ &
           &
           & 
    $0.95$ &
    $1.06$ \\
    $^{12}$C &
    $0.500$ &
    $16.08$ &
    $16.06$ &
    $28.96$ &
    $28.92$ &
   $32.4$ &
   $22.38$ &
    $1.00$ &
    $1.00$ \\
    $^{16}$O &
    $0.500$ &
    $15.61$ &
    $15.59$ &
    $29.48$ &
    $29.43$ &
    $30.9$ & 
    $23.81$ &
    $1.00$ &
    $1.00$ \\
    $^{27}$Al &
    $0.481$ &
    $16.61$ &
    $16.92$ &
    $30.93$ &
    $30.26$ &
            & 
    $25.12$ &
    $0.98$ &
    $1.02$ \\
    $^{40}$Ca &
    $0.500$ &
    $16.44$ &
    $16.42$ &
    $31.23$ &
    $31.18$ &
    $33.8$ &
    $27.72$ &
    $1.00$ &
    $1.00$ \\
    $^{48}$Ca &
    $0.417$ &
    $15.64$ &
    $17.84$ &
    $33.04$ &
    $30.06$ &
            & 
    $27.05$ &
    $0.88$ &
    $1.10$  \\
    $^{56}$Fe &
    $0.464$ &
    $16.71$ &
    $17.45$ &
    $32.33$ &
    $31.13$ &
    $32.7$ & 
           &
    $0.96$ &
    $1.04$ \\
    $^{108}$Ag &
    $0.435$ &
    $16.48$ &
    $17.81$ &
    $33.55$ &
    $31.16$ &
           &
           & 
    $0.93$ &
    $1.08$ \\
\br
\end{tabular}
\end{indented}
\end{table}

\begin{figure}[tpb] 
\begin{center}
        \includegraphics[width=0.5\textwidth]{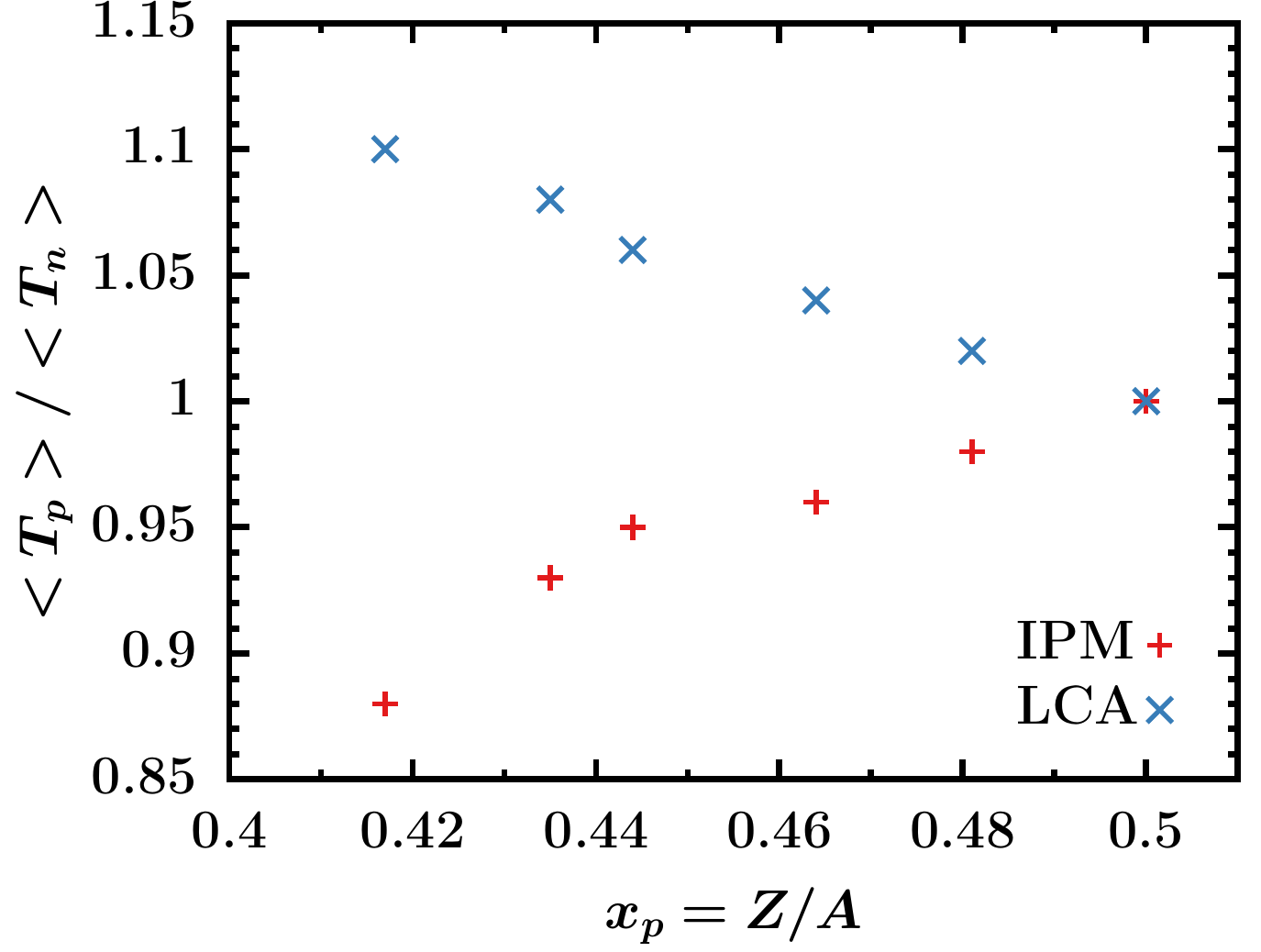}
\end{center}
\caption[]{ The IPM and LCA predictions for the $\left< T_{p} \right>
  / \left< T_{n} \right>$ as a function of the proton fraction
  $x_{p}$.
      \label{fig:xpdepen}
      }
\end{figure}

Table~\ref{tab:kinenergy} compares the IPM and LCA predictions for the
kinetic energies per proton and neutron. Obviously, as the kinetic
energies can be associated with the fourth moments of the
$n^{[1]}(p)$, they are highly sensitive to its fat tails. Indeed,
inclusion of the correlations increases the $\left< T_{p} \right>$ and
$\left< T_{n} \right>$ by a factor of about two. For the sake of reference, the average kinetic
energy of a one-component nuclear Fermi gas is 21~MeV. For the
heaviest nuclei studied in this work we find values which are about
50\% larger. The LCA results for the average kinetic energies for
$^{9}$Be are comparable to those of realistic calculations
quoted in Table~1 of~\cite{Sargsian:2013gea} --- $\left< T_{p}
\right>=29.82$~MeV and $\left< T_{n} \right>$=27.09~MeV. 
As can be appreciated from Table~\ref{tab:kinenergy}, the LCA
predictions for the correlated kinetic energies $\left< T_{N} \right>$
are comparable with those of the realistic model of
\cite{Claudio1996}. The predictions for $\left< T_{N} \right>$ from
the variational calculations of~\cite{Feldmeier:1997zh} are
systematically smaller. The $^{16}$O kinetic energies reported in
\cite{Pieper:1992gr} --obtained using the same spin-isospin and tensor
correlation functions as used here--, are about 10\% larger than those
from the LCA approach (33.7~MeV and 33.8~MeV, compared to 29.4~MeV and
29.5~MeV). Part of this discrepancy can be attributed to the fact that
we have imposed a hard momentum cutoff at $4.5$~\invfm and this is not
present in the calculations of \cite{Pieper:1992gr}. Incrementing the
hard momentum cutoff to $6.0$~\invfm, which completely invalidates the use of a non-relativistic framework, increases the LCA predictions for the $\left< T_{N} \right>$ with about 15\%.

The parameter $x_p= \frac{Z}{A}$ is the proton fraction and is a measure for
the asymmetry of nuclei. As expected for a non-interacting two-component Fermi
system, $\left< T_{p} \right>< \left< T_{n} \right>$ for asymmetric nuclei ($x_{p}<0.5$) in the IPM.
As can be appreciated from figure~\ref{fig:xpdepen} after inclusion of the
correlations, the situation is reversed with the minority component having a
larger average  kinetic energy.
This can be attributed to the tensor correlations,
which are stronger between $pn$ than between $pp$ and $nn$ pairs.
The difference between $\left< T_{p} \right>$ and $\left< T_{n} \right>$
increases roughly linearly with decreasing proton fraction $x_p$. For the
most asymmetric nucleus considered here, $^{48}$Ca, $\left< T_{p} \right>$
is about 10\% larger than $\left< T_{n} \right>$.

We now discuss the effect of the correlations on the root-mean-square
(rms) radii of the nuclear matter distribution. The rms radii can be computed with an operator of the form
\begin{equation}
{\widehat{r^2}}= \frac{1}{A} \sum _{i} \left( \vec{r}_i - \vec{R}_{cm} \right)^2 ,
\end{equation} 
with $\vec{R}_{cm}=\frac{1}{A} \sum_{i} \vec{r}_i$. Using a procedure
which is completely similar to the one used for the kinetic energy, in
the LCA the operator ${\widehat{r^2}}$ becomes a correlated operator
with a one-body and a two-body part.  Table~\ref{tab:rmsradii}
compares the IPM and the LCA predictions for the rms radii. The IPM
predictions which are obtained with the global parametrisation
of~(\ref{eq:paramho}) tend to overestimate the measured radii for
light and heavy nuclei, and underestimate them for mid-heavy
nucleus. All in all, the effect of the correlations on the computed
rms radii is rather modest. Inclusion of the correlations reduces the
rms radii bij 8-12\%. The reduction factor is hardly dependent on the
atomic mass number. Central correlations introduce an exclusion zone
around each nucleon and are therefore expected to increase the
computed rms radii. The dominant role of the other correlations and
the effect of the normalization (which is reflected in the
$\frac{1}{\mathcal{N}}$ factor in the effective operator) are at the
origin of the modest reduction for the rms radii after including the
SRC in the LCA framework.  The LCA predictions for the rms radii are
in acceptable agreement with the experimental values and the
predictions from the UCOM framework of~\cite{Feldmeier:1997zh}. We
stress that our IPM results are obtained with a single Slater
determinant with HO wave functions from the global parametrisation of
equation~(\ref{eq:paramho}). It is likely that one can find a slightly
modified parametrisation that brings the LCA rms radii closer to the
data.

\begin{table}[tb]
  \caption{Results from the IPM and LCA framework for the rms radii
    for a variety of nuclei. The results are compared with those from the Unitary Correlation Operator Method (UCOM) \cite{Feldmeier:1997zh} and experimental values (Expt) \cite{Angeli2013}. All radii are in fm. 
  \label{tab:rmsradii}
}
\begin{indented}
\item[]\begin{tabular}{@{}ccccc}
\br
 & & & \\
  $A$ & IPM & LCA  & UCOM  
\cite{Feldmeier:1997zh}  & Expt \cite{Angeli2013} \\
\mr
$^4$He & 1.84 & 1.70 & 1.35 & 1.6755 $\pm$ 0.0028\\
$^{9}$Be & 2.32 & 2.13 & & 2.5190 $\pm$ 0.0120 \\
$^{12}$C & 2.46 & 2.23 & 2.36 & 2.4702 $\pm$ 0.0022 \\
$^{16}$O & 2.59 & 2.32 & 2.28 & 2.6991 $\pm$ 0.0052 \\
$^{27}$Al & 3.06 & 2.72 & 2.82 & 3.0610 $\pm$ 0:0031 \\
$^{40}$Ca & 3.21 & 2.84 & 2.93 & 3.4776 $\pm$ 0.0019 \\
$^{48}$Ca & 3.47 & 3.05 & 3.20 & 3.4771 $\pm$ 0.0020 \\
$^{56}$Fe & 3.63 & 3.20 & & 3.7377 $\pm$ 0:0016 \\
$^{108}$Ag & 4.50 & 3.94 & & 4.6538 $\pm$ 0.0025\\
$^{197}$Au &  5.73 & 5.21 & & 5.4371 $\pm$ 0.0038 \\
$^{208}$Pb & 5.83 & 5.28 & & 5.5012 $\pm$ 0.0013 \\
\br 
\end{tabular}
\end{indented}
\end{table}

\section{Summary}

We have introduced an approximate method, dubbed LCA, for the
computation of the SRC contributions to the single-nucleon momentum
distributions $n^{[1]}(p)$ throughout the whole mass table. A basis of
single-particle wave functions and a set of correlation functions
serves as an input to LCA. For the numerical calculations presented
here, we have included the central, spin-isospin and tensor
correlations and mass-independent correlation functions.  The LCA
method predicts the characteristic high-momentum part of the
single-nucleon momentum distribution for a wide range of nuclei.  For
the light nuclei $^4$He, $^9$Be and $^{12}$C, the LCA predictions for
the tails of the single-nucleon momentum distributions reproduce the
stylized features of the QMC ones obtained with realistic
Hamiltonians. The predicted aggregated effect of SRC and its mass
dependence closely matches the observations from inclusive electron
scattering ($a_2$ coefficients and the magnitude of the EMC effect).

In the LCA, one can separate contributions of the central,
spin-isospin and tensor correlations and study how these affect the
relative strength of $nn$, $pp$ and $pn$ pairs in the high-momentum
tail of $n^{[1]}(p)$.  For $1.5 \lesssim p \lesssim 3$~\invfm the
$n^{[1]}(p)$ is dominated by tensor-induced $pn$ correlations.  Our
prediction for the relative strength of $pp$ and $pn$ pairs in the
tail part of $n^{[1]}(p)$ is in line with observations in exclusive
two-nucleon knockout studies which point at a strong dominance of $np$
SRC pairs over the $pp$ SRC pairs.  We have shown that the
high-momentum tail of $n^{[1]}(p)$ is dominated by the correlation
operators acting on mean-field pairs with vanishing relative radial
quantum number and vanishing orbital angular momentum, i.e.~IPM pairs
in a close-proximity configuration.  Another prediction of the LCA is
that in asymmetric nuclei, the correlations are responsible for the
fact that the average kinetic energy of the minority nucleons is
larger than for the majority nucleons. The LCA method provides results
for the correlated average kinetic energies and nuclear radii which
are in line with those of alternate approaches.

\ack
We thank O.~Hen and E.~Piasetzky for many fruitful discussions. 
This work is supported by the Research Foundation Flanders (FWO-Flanders) and
by the Interuniversity Attraction Poles
Programme P7/12 initiated by the Belgian Science Policy
Office. The computational resources (Stevin Supercomputer Infrastructure) and
services used in this work were provided by Ghent University, the Hercules
Foundation and the Flemish Government.

\section*{References}

\end{document}